\documentclass{pazhb}

\usepackage[hyperfootnotes=false]{hyperref}

\usepackage{color}
\usepackage{epsf}

\usepackage[utf8]{inputenc}
\usepackage[english]{babel}
\usepackage[T2A]{fontenc}

\usepackage{amsmath}
\usepackage{amsfonts}
\usepackage{amssymb}
\usepackage{multirow}

\usepackage{epsf}
\usepackage{graphicx}
\usepackage{gensymb}
\usepackage{float}
\usepackage{longtable}
\usepackage{changes}
\DeclareGraphicsExtensions{.pdf,.png,.jpg}
\renewcommand{\AA}{\ensuremath{\mathring{A}}}

\usepackage{subcaption}
\def\beq#1{\begin{equation}\label{#1}}
\def\eeq{\end{equation}}
\def\beqa#1{\begin{eqnarray}\label{#1}}
\def\eeqa{\end{eqnarray}}

\def\comment#1{\relax}

\def\apgt{\ {\raise-.5ex\hbox{$\buildrel>\over\sim$}}\ }
\def\aplt{\ {\raise-.5ex\hbox{$\buildrel<\over\sim$}}\ }

\newcommand{\kms}{km\,s$^{-1}$}

\begin{document}

\title{Radial velocities of narrow emission line components in the spectra of T~Tauri stars}

\author{
  V.~A.~Kiryukhina\email{valeriya@sai.msu.ru}\address{1},
  A.~V.~Dodin \address{1}
  \addresstext{1}{Sternberg Astronomical Institute of Moscow State University} \\
}

\shortauthor{V.A. Kiryukhina and A.V. Dodin}

\shorttitle{Radial velocities of narrow emission lines in the spectra of cTTs}

\begin{abstract}
We studied rotational modulation of the radial velocities of narrow emission lines in four classical T Tauri stars. We found that the previously declared shift of the mean velocity of neutral and ionized helium lines relative to the mean radial velocity of the star is not associated with the inflow of accreted gas into the hotspot, since the radial velocity curves for lines with different velocity shifts should exhibit phase shifts relative to each other, while the observed phase shifts are absent within their uncertainties and do not correspond to the observed line velocity shifts. This means that the line shifts are not caused by the actual gas motion. For neutral helium lines, the shifts can be explained by the large optical thickness of the lines and the Stark effect at plasma parameters expected at the base of the accretion column of T Tauri stars.

\keywords{accretion, accretion discs, shock waves, stars: variables: T Tauri }

\end{abstract}


\section*{Introduction}
The activity of T Tauri stars is attributable to the magnetospheric accretion of matter from a protoplanetary disk \citep{2016ARA&A..54..135H}. As the gas penetrates the magnetosphere, it travels along magnetic field lines and decelerates at the shock front in the stellar atmosphere. 
The gas heated at the shock front cools and settles into the deeper layers of the atmosphere, gradually mixing with stellar matter. The radiation from the hot post-shock irradiates and heats the colder regions of the post-shock. As the gas cools, the importance of this heating increases until a balance between cooling and heating is achieved. This region, where thermal balance is established, is commonly called as a hotspot \citep{2018MNRAS.475.4367D}.

The hotspot primarily manifests as an additional continuum, reducing the contrast of absorption lines (the veiling effect), and as narrow emission lines of HeI, HeII, neutral, and singly ionized metals. Numerous measurements of the radial velocity of these lines show sinusoidal variations caused by the motion of the hotspot as the star rotates on its axis. In this model, the amplitude of the radial velocity curve is determined by the stellar rotation velocity $v \sin i$ and the location of the spot on its surface, allowing the latitude of the accretion zone to be estimated (\citealt{2020MNRAS.497.2142M}).
At the same time, in the work \citet{2020MNRAS.497.2142M} and in numerous other studies (\citealt{2001A&A...369..993P, 
2021A&A...656A..50P, 2023A&A...670A.165N, 2024arXiv240903322P}  and others) it was found that the oscillations of the radial velocity of the emission lines do not occur around the stellar radial velocity, but around a value shifted by several {\kms} to the red side. This is interpreted as residual gas settling in the stellar atmosphere. The presence of such a settling with a velocity of several {\kms} contradicts the physics of the shock. In the case of a stationary shock from equations of conservation of mass and momentum:
\begin{equation}
\begin{cases}
    \rho_{\rm in} V_{\rm in}=  \rho V, \\
    \rho_{\rm in} V_{\rm in}^2 +P_{\rm in}=  \rho V^2 +P,
    \label{eq:conserv}
\end{cases}
\end{equation}
for an ideal gas with $P_{\rm in}\ll\rho_{\rm in} V_{\rm in}^2$ temperature $T$ and velocity $V$ are related by the ratio:
\begin{equation}
    T = \frac{(V_{\rm in}-V)\mu m_p V}{k},\label{eq:Temp}
\end{equation}
here $\rho_{\rm in},$ $V_{\rm in},$ $P_{\rm in}$ are pre-shock values of density, velocity, and pressure of the infalling gas, and $\rho,$ $V,$ $P$ are the corresponding post-shock values at an arbitrary point, $\mu=0.6$ -- mean molecular weight of ionized gas, $m_p = 1.67\times 10^{-24}${\it g}, $k = 1.38 \times 10^{-16}$ erg\,K$^{-1}$.
For a typical fall velocity $V_{\rm in} = 300$ km\,s$^{-1}$, the measured velocity of the helium lines $V \sim 10$ km\,s$^{-1}$ corresponds to a temperature $\sim 2\times10^5$ K. 

However, neutral helium cannot exist at a such temperature, and the He\,I line cannot be shifted at the temperature where neutral helium forms, since, according to Eq.\,\ref{eq:Temp} a temperature of $2\times10^4$ К corresponds to a velocity $\simeq 1$ km\,s$^{-1}$.

More detailed calculations \citep{2018MNRAS.475.4367D} reveal that the line formation region shifts to deeper and slower layers of the hotspot due to delayed recombination effects and ionizing radiation from the upper layers. Thus, the current measurements of the radial velocities of the narrow components of the helium lines contradict the predictions of the steady-state shock model.

The question of whether the observed line shifts indicate the non-stationarity of the shock, which should arise as a result of thermal instability (about the shock instability see \citealt{2016A&A...594A..93C}), is complex, as both photoionization and delayed recombination also occur in the non-stationary case. These factors shift the region of helium line formation from the post-shock, where thermal instability develops, to the hotspot, where the gas has lost its velocity ($V<1$ \kms), reached radiative equilibrium, and where line formation occurs due to irradiation from the overlying layers. 
However, detailed calculations of helium line formation for an unstable shock have not been conducted, and it cannot be ruled out that the observed shifts are a result of this instability. Otherwise, we must acknowledge that we do not understand the origin of these lines or why they are redshifted.

On the other hand, there are doubts about the reliability of the radial velocity measurements of the helium lines. Conclusions regarding the shift of the narrow components are based on the He\,I 5876 and He\,II 4686\,{\AA} lines. These are multicomponent lines, and it is unclear what the authors consider as the central wavelength. The velocity of the single-component line He\,I 6678\,{\AA} has not been measured, likely because it is superimposed by the stellar absorption line. In addition, the lines are formed in dense gas, and the Stark effect can also shift the neutral helium lines to both the red and blue sides of the spectrum. These complexities and uncertainties prompted us to conduct our own study of the radial velocities of narrow emission components, using data from the longest spectral monitoring available in the archives. In addition to the narrow component, the profiles of the lines contain a broad component that is not associated with the hotspot. To mitigate the influence of this broad component on our results, we restricted our study to four stars: BP Tau, DK Tau, EX Lup, and TW Hya, where the narrow components dominate the profiles of at least several lines.

\section{Observations}\label{sect:obs}
We obtained the spectral monitoring data for the four stars from the archives of the CFHT\footnote{\url{www.cadc-ccda.hia-iha.nrc-cnrc.gc.ca}} and ESO\footnote{\url{https://archive.eso.org/scienceportal/home}} observatories. The number of spectra, observation dates and spectrographs we used for each star are listed in Table \,\ref{tab:observations}. In addition to the stars we study, we need a template spectrum, namely the spectrum of the weak-lined T Tauri star TAP\,45. The main characteristics of the spectrographs we used and details on additional spectral processing are below.

\begin{enumerate}
    \item ESPaDOnS (CFHT).
The spectral resolution is $\sim 68\,000$ within the range of 3700 to 10480 {\AA}. All spectra were obtained from the data archive in a calibrated, one-dimensional format for each spectral order. In the overlapping regions of the orders, the profiles of the spectral lines coincide within the noise limits. Therefore, to simplify the analysis, we combined the data from all orders into a single array and sorted it by wavelength. Thus, in the overlapping regions of the spectral orders, there is an overlay of two wavelength grids. We did not convert the spectrum to a single grid to preserve the statistical properties of the original noise. Additionally, all spectra were normalized to the continuum. The data have varying signal-to-noise ratios (SNR) ranging from 11 to 75.

\item FEROS (MPG/ESO). The spectral resolution is $\sim 48\,000,$ within the range of 3500 to 9200 {\AA}. SNR ranges from 14 to 55.

We excluded spectra with low SNR < 20 and those with numerous outliers caused by cosmic particles. Therefore, our dataset includes 32 spectra collected over 22 nights. Spectra taken consecutively during a single night showed no changes and were averaged.

\end{enumerate}

\begin{table}
\caption{Observations}
\label{tab:observations}
\centering
\begin{tabular}{c c c}
\hline\hline
Spectrograph & $\rm N_{spec}$  & Dates \\
\hline
\multicolumn{3}{c}{BP\,Tau} \\
 ESPaDOnS  &  9   & 01.11.11 -- 15.11.11  \\
           &  13  & 04.01.12 -- 17.01.12  \\
           &  11  & 08.01.14 -- 21.01.14  \\
\hline
\multicolumn{3}{c}{DK\,Tau}\\
 ESPaDOnS  &  7  & 15.12.10 -- 30.12.10  \\
           &  9  & 25.11.12 -- 23.12.12  \\
           &  5  & 04.02.17 -- 16.02.17  \\
\hline
\multicolumn{3}{c}{EX\,Lup}\\
 FEROS    &  9  & 03.07.12 -- 13.07.12 \\
          &  6  & 15.07.13 -- 27.07.13 \\
          &  3  & 11.09.13 -- 15.09.13 \\
          & 14  & 16.02.14 -- 21.02.14 \\
 ESPaDOnS & 11  & 09.06.16 -- 24.06.16 \\
          &  6  & 31.05.19 -- 12.06.19 \\
\hline
\multicolumn{3}{c}{TW\,Hya}\\
 ESPaDOnS &  13 & 15.03.08 -- 28.03.08 \\
          &  13 & 23.02.10 -- 08.03.10 \\
          &  13 & 31.01.12 -- 13.02.12 \\
          &  15 & 17.02.16 -- 02.03.16 \\
\hline
\multicolumn{3}{c}{TAP\,45}\\
 ESPaDOnS & 13 &           \\
\hline
\end{tabular}
\end{table}

\section{Measurement of radial velocities of absorption lines}

To measure the velocities of helium lines in the stellar reference frame, we need to know the stellar velocity, which can be derived from the absorption lines. However, in the case of cool stars, the lines often blended, and even slight overlap can cause a significant shift in the line center. To reduce this effect, we can use a synthetic spectrum that closely fits the observations. Instead of a synthetic spectrum, a high-quality observed spectrum of a similar non-accreting star can also be used. We found that the latter approach gives the most accurate results when using the star TAP45 as the template spectrum. \cite{2023A&A...670A.165N} used this star to measure the veiling of DK Tau. Since DK Tau and BP Tau have similar parameters, we used TAP45 as a template spectrum for both stars to measure radial velocity variability and veiling.

\citet{2023A&A...670A.165N} reported $v \sin i=11.5$ km\,s$^{-1}$ for TAP\,45, however, our calculations using pySME\footnote{\url{https://pysme-astro.readthedocs.io/en/latest/}} (the Python version of Spectroscopy Made Easy) (\citealt{2023A&A...671A.171W}, \citealt{2017A&A...597A..16P}, \citealt{1996A&AS..118..595V}) revealed that at this velocity, the theoretical profiles are marginally broader than the observed ones. For the best fit of the profiles, we calculated the spectra using different values of $v \sin i$ and found that the optimal fit in the range of 5500 -- 9000 {\AA} is achieved at $v \sin i = 7$ km\,s$^{-1}$, with simultaneous broadening by a Gaussian profile with $\sigma=2.5$ \kms.

To use the TAP45 spectrum as a template for an absorption spectrum, we must adjust the line widths. DK Tau and BP Tau have larger $v \sin i$ values than TAP\,45, so we must broaden the spectrum of TAP\,45. Generally, this broadening should be performed by convolving the spectrum with a rotational profile. However, in practice, the profiles are adequately reproduced by convolving with a Gaussian profile with $\sigma=6.7$ {\kms} for DK Tau and $\sigma=3.4$ {\kms} for BP Tau.

EX Lup and TW Hya exhibit lower rotational velocities compared to TAP\,45 (Table.\,\ref{tab:starpar}). Consequently, we determined the radial velocities of EX Lup and TW Hya relative to a synthetic spectrum, using the pySME program. To specify model parameters for the calculations, we adopted an effective temperature of $T_{\rm eff} = 3750$ K for EX Lup and $T_{\rm eff} = 3810$ K for TW Hya(\citealt{2014A&A...561A..61K}, \citealt{2023ApJ...956..102H}), with $\log{g}= 4.0$ and solar metallicity for both stars. The list of lines was obtained from the VALD database \citep{2015PhyS...90e4005R}.

\begin{table*}
\caption{Stellar parameters}
\label{tab:starpar}
\centering
\begin{tabular}{cccccc}
\hline\hline
Star  &  Spectral & $v_r$,           & $v\,sin\,i$, & Period, & $i, $\\
        & type &  \kms            & \kms         & d. & $\degree$\\
\hline
BP\,Tau & M0.5$^{a}$ & 15.8 $\pm$ 0.3   &  9$^{a}$   & --     & 38$^{b}$ \\
DK\,Tau & K7$^{c}$ & 16.1 $\pm$ 0.3   & 13$^{c}$   & 9.390 & 58$^{c}$ \\
EX\,Lup & M0$^{d}$ & -0.38 $\pm$ 0.06 &  4.4$^{d}$ & 7.417 & 32$^{e}$ \\
TW\,Hya & M0.5$^{f}$ & 12.67 $\pm$ 0.02 &  4$^{g}$   & 3.568 & 15$^{g}$ \\
TAP\,45 & K6$^{f}$ & 18.42 $\pm$ 0.02 &  7   &   --  &  -- \\
\hline
\multicolumn{6}{l}{ $^b$\cite{2008MNRAS.386.1234D}, $^b$ \cite{2019ApJ...882...49L} , $^c$\cite{2023A&A...670A.165N},} \\
\multicolumn{6}{l}{$^d$\cite{2015A&A...580A..82S}, $^e$\cite{2024arXiv240405420S},} \\
\multicolumn{6}{l}{$^f$\cite{2014ApJ...786...97H}, $^g$\cite{2011MNRAS.417..472D}.}
\end{tabular}
\end{table*}

To determine the radial velocities of the stars, we selected intervals containing isolated absorption lines with insignificant blending and fitted the theoretical spectrum $f_{\rm T}$ to the observed spectrum $f_{\rm C}$:
\begin{equation}
{\left\|f_{\rm C} - C_i\frac{f_{\rm T}[\lambda(1+\frac{v_{i}}{c})]+r_i}{(1+r_{i})}\right\| \rightarrow {\rm min},}
\end{equation}
where the unknown parameters $v_i$, $r_i$ and $C_i$ are the velocity, veiling, and the factor correcting the continuum level in the i-th wavelength interval are determined using the least squares method.

Thus, weighted average values of radial velocities $v(t_i)$ were obtained for each individual spectrum of all the studied stars. In the case of BP Tau, non-periodic variability in $v(t)$ was observed. Consequently, the median value of the radial velocity over all observations was calculated to be $v_ r = 15.8 \pm 0.3$ \kms. For DK Tau, EX Lup, and TW Hya, the weighted average radial velocities $v(t_i)$ exhibit the periodic variability that can be approximated by a sinusoid:

\begin{equation}
    v_i = A \cos \frac{2\pi(t_i - t_0)}{P}+ \overline{v} \label{cos},
\end{equation}
where $v_i$ is the weighted average velocity of the i-th spectrum, $t_i$ is the Julian date of the observation, P is the period in days, and $\overline v $ is the average velocity. The parameters are determined using the least squares method, with weights inversely proportional to the error in $v_i$: $w_i=\sigma_{i}^{-2}$. The resulting average radial velocities of the stars are summarized in the Table.\,\ref{tab:starpar}.

\section{Measurement of radial velocities of emission lines}
\subsection{He\,I, He\,II}\label{sect:HeI}

To measure the velocities of neutral and ionized helium lines, it is essential to first determine the central wavelengths of these lines. Although this question may seem trivial, different authors often use varying wavelengths, typically without specifying the value with adequate precision. For example, in \cite{2024arXiv240405420S}, the laboratory value from NIST for the He\,I 5876 line is used: 5875.621 {\AA}. In contrast, the average theoretical $gf$ value for this line is 5875.66 {\AA}, which differs by 2 \kms. In this work, we adopt the wavelength averaged with the weight of the oscillator strength $gf$ of each component as the central wavelength for multicomponent lines:
\begin{equation}
    \lambda_0 = \frac{\sum{\lambda_i{(gf)_i}}}{\sum{(gf)_i}},\label{eqgf}
\end{equation}
which is valid for an optically thin line in the low-density limit. The line shift caused by not satisfying these conditions will be considered relative to this $\lambda_0$. The wavelengths and oscillator strengths of each line component are obtained from the NIST database.

The second issue in measuring the radial velocities of emission lines is their blending with absorption lines. Specifically, before measuring the velocity of the He\,I 6678\,{\AA} line, it is necessary to remove the Fe\,I 6677.9 {\AA} absorption line, which is superimposed on the blue side of the He\,I profile. The profile of the ionized helium line He\,II 4686\,{\AA} is also blended by several absorption lines (the strongest lines -- Ca\,I 4685.3, Ni\,I 4686.2\,{\AA})

\begin{figure}
    \centering
    \includegraphics[width=\columnwidth]{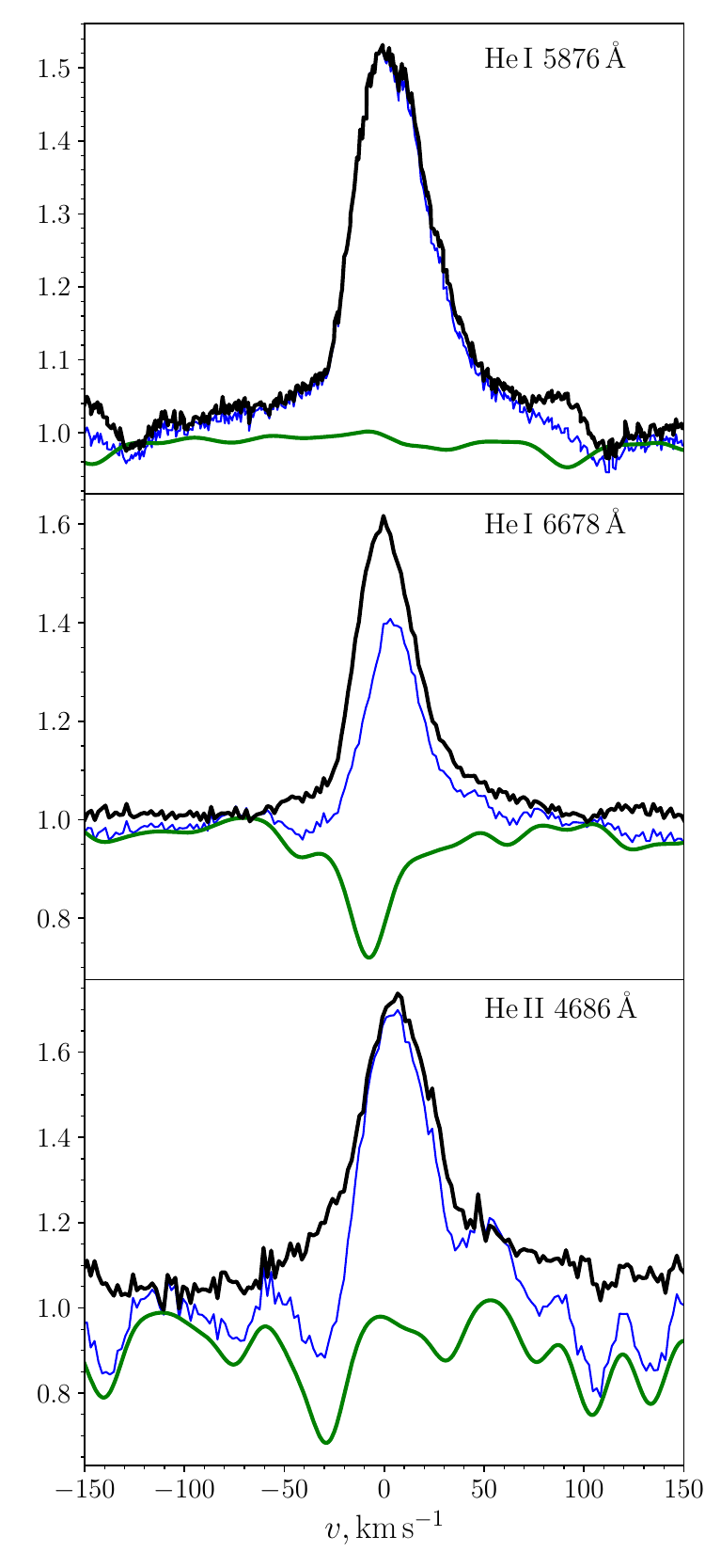}
    \caption{Removing the stellar absorption lines for DK Tau. The blue lines represent the observed He\,I 6678\,{\AA} and He\,II 4686\,{\AA} lines; the green lines are the TAP\,45 spectrum, scaled according to the veiling effect; the black lines show the corrected He\,I 6678\,{\AA} and He\,II 4686\,{\AA}, lines without absorption lines; For He\,I 5876\,{\AA}, the observed profile is shown. The observation date is MJD = 56262.443.}
    \label{fig:profiles}
\end{figure}

In general, the spectrum of a star with an accretion spot is not a simple sum of the spectra of the star and the spot, but must be derived by integrating the intensities over the surface of the star. The separation of the spectra of a star and a spot is a complex and still unsolved problem. To reduce the influence of absorption lines on the obtained results, we approximate the absorption spectrum of the star using the TAP\,45 spectrum (for DK\,Tau and BP\,Tau) or the synthetic spectrum (for EX\,Lup and TW\,Hya) with the required veiling parameters. The veiling value was determined based on the region near the He\,I 6678\,{\AA} and He\,II 4686\,{\AA} lines. After subtracting the absorption spectrum, we also required that the profile of the broad component of the He\,I 6678\,{\AA} line match the profile of the broad component of the He\,I 5876\,{\AA} line. Examples of initial profiles and subtraction results are shown in Fig.\,\ref{fig:profiles}.

The neutral helium lines consist of narrow and broad components, which originate from distinct spatial regions with various physical conditions. Consequently, the ratio between these components can differ for various lines. However, the shape of these components must be the same. These two line components can be approximated as the sum of two Gaussians, each characterized by its own set of parameters. The profiles of the He\,I 5876\,{\AA} and He\,I 6678\,{\AA} lines can be described simultaneously using the least squares method. The function representing the two components for these lines can be expressed as a sum of four Gaussian functions:
\begin{multline}
{y = a_1g(v_1,\sigma_1) + a_2g(v_2,\sigma_2) + }\\
{ a_3g(v_3,\sigma_3) + a_4g(v_4,\sigma_4) + c,}
\end{multline}
where $g(v,\sigma)$ is the Gaussian function, $c$ is the continuum level. The velocities $v_1$ and $v_3$ correspond to the narrow components of the He\,I 5876\,{\AA} and He\,I 6678\,{\AA} lines, respectively. These velocities are treated as different due to the Stark effect and the multicomponent nature of the He\,I 5876\,{\AA} line. The velocities and widths of the broad components are assumed to be identical $v_2 = v_4,  \sigma_2 = \sigma_4$ to achieve a more stable solution, particularly for the weaker He\,I 6678{\AA} line. The neutral helium velocities of DK Tau and TW Hya were calculated in this way. But in the spectra of EX Lup, the broad component of the He\,I lines is weakly expressed, therefore, the neutral helium line profiles were described separately by selecting the parameters of a single Gaussian for each line.

To calculate the velocity of the He\,II 4686 {\AA} line in the absorption-corrected spectra of all stars, the line profile was described as the sum of two Gaussians. The center of the Gaussian representing the narrow component was taken as a center of the He\,II line.

In the BP Tau spectra, the Gaussian function does not adequately describe the shape of the helium line profiles. Therefore, to measure the velocity of the narrow line components, their `center of gravity' was calculated using the following formula:
\begin{equation}
    \lambda_c = \frac{\int(1-f_{\rm C})\lambda \,\mathrm{d}\lambda}{{\rm EW}} , \label{eqlc}
\end{equation}
where ${\rm EW}$ is the equivalent line width, $f_{\rm C}$ is the continuum-normalized spectrum. To minimize the contribution of broad components, only the upper 60\% of the profile was considered.

To estimate the error introduced by the inaccuracy of subtracting the stellar absorption spectrum from the measured velocities of He\,I 6678\,{\AA} and He\,II 4686 {\AA}, we measured the line center at the permissible upper ($r_{max}$) and lower ($r_{min}$) limits for veiling, in addition to measurements for the optimal veiling value ($r$). The differences between the velocity values obtained at different veiling levels are included in the measurement error. The final results of measuring the velocities of neutral and ionized helium in the spectra of the stars are presented in the Table\,\ref{tab:vHe}. The obtained velocities deviate from zero, potentially indicating the presence of gas movement in the atmosphere.

\begin{table}
\caption{Velocities of He\,I, He\,II lines and average velocity of metal lines.}
\label{tab:vHe}
\centering
\begin{tabular}{c l r}
\hline\hline
 Star & Line  & \multicolumn{1}{c}{$\overline{v},$ \kms} \\
\hline
BP Tau & He\,I 5876  &  $4.0 \pm 0.1 $ \\
& He\,I 6678  & $1.9 \pm 0.3 $ \\
& He\,II 4686  & $7.6 \pm 0.1 $ \\
& Metals     &  $-0.6 \pm 0.8 $  \\
\hline
DK Tau & He\,I 5876 &  $4.3 \pm 0.3 $ \\
& He\,I 6678  &  $2.4 \pm 0.5 $ \\
& He\,II 4686  &    $6.9 \pm 0.6 $ \\
& Metals     &    $-1.0 \pm 0.5 $  \\
\hline
EX Lup & He\,I 5876  &  $3.5 \pm 0.2 $ \\
& He\,I 6678  & $2.1 \pm 0.4$   \\
& He\,II 4686  & $6.9 \pm 0.4$    \\
& Metals     &  $0.5 \pm 0.2$  \\
\hline
TW Hya & He\,I 5876  &  $4.1 \pm 0.1$     \\
& He\,I 6678  & $4.2 \pm 0.2$   \\
& He\,II 4686  &  $6.5 \pm 0.2$   \\
& Metals     &  $0.3 \pm 0.1$ \\
\hline
\end{tabular}
\end{table}

\subsection{Radial velocities of metal lines}
The emission lines of metals are formed in regions deeper than the helium lines \citep{2018MNRAS.475.4367D}, and thus their expected velocity relative to the stellar surface should be approximately zero. To verify this hypothesis and validate our methods, we measured the six strongest narrow emission lines of metals (Fe\,II 4924\,{\AA}, Fe\,II 5018\,{\AA}, Fe\,II 5170\,{\AA}, Fe\,II 5234\,{\AA}, Mg\,I 5172\,{\AA},  Mg\,I 5183\,{\AA}) in the spectra of the studied stars and calculated their weighted average for each spectrum:
\begin{equation}
    \overline{v}_t = \frac{\sum{v_i w_i}}{\sum{w_i}}, \quad w_i = \frac{1}{\sigma^2_i} \label{wav}
\end{equation}
The results are presented in the Table\,\ref{tab:vHe}.

\subsection{Periodicity of radial velocities} \label{sect:BP}
The primary cause of the variations in the radial velocities of the narrow emission lines is the change in the radial velocity of the hotspot due to the stellar rotation. The rotation periods of the stars TW Hya and EX Lup can be determined from the sinusoidal variations in the radial velocity of the absorption lines. The periods obtained using this approach are 3.568 days for TW Hya and 7.417 days for EX Lup, which are consistent with previous studies (\citealt{2023ApJ...956..102H}, \citealt{2014A&A...561A..61K}).

The points on the radial velocity curves from absorption lines for DK Tau and BP Tau exhibit significant deviations from the expected sinusoidal curve. Furthermore, in the same spectrum, different lines display different shifts. These deviations are not attributable to measurement errors but are of natural origin, associated with the veiling effect. As a result of the stellar rotation, any deviation from axial symmetry results in variable profile distortions, leading to apparent variability in radial velocities. Strict periodicity is also not evident in the photometric curve due to the superposition of other variability factors (accretion variability, dust eclipses).

The periodicities in the variability of the emission lines were studied using the Barning method \citep{1992T} for all measured emission lines within the interval of 3 to 15 days. The periodogram does not display a clearly defined period, with the maximum peak for DK Tau at P = 9.3902 days. However, the periods in the range of $\sim 8.2 - 8.4$ days reported in the literature \citep{2023A&A...670A.165N, 1993A&A...272..176B, 2010PASP..122..753P} also cannot be dismissed; for instance, a peak is observed at the period P = 8.1887 days.

In the case of BP Tau, strict periodic variability in the emission lines is also absent. Regular changes in radial velocity occur over short time intervals, leading to the conclusion that the rotation period of BP Tau is $\sim 8$ days, which is within the typical range for T Tauri stars.

In addition, an interesting feature of BP Tau is that the variation in the velocity of the emission lines ($V_{\rm e}$), unlike all other stars, occurs in phase with the variation in the velocity of the absorption lines ($V_{\rm a}$). It is impossible if the cause of both changes is a variation in the radial velocity of the hotspot due to stellar rotation \citep{2001A&A...369..993P}. The coincidence of the phases of $V_{\rm e}$ and $V_{\rm a}$ would be observed, if the matter flows into the hotspot tangentially to the surface. However, this does not align with the generally accepted picture of the formation of narrow components of emission lines. Distortions in the radial velocity of the stellar absorption lines can be caused by inhomogeneous eclipses of the stellar disk, as a consequence of the Rossiter–McLaughlin effect \citep{2021MNRAS.503.5704D}. If such eclipses occur with a rotation period, they can cause a phase shift in the radial velocity by an arbitrary amount. The cause of such eclipses may be the deformation of the inner boundary of the accretion disk \citep{1999A&A...349..619B}, which  eclipses the star periodically.

\section{The radial velocity curve in the presence of radial gas motion.} \label{sect:Deltaphi}

\begin{figure}
    \centering
   \includegraphics[width=0.75\columnwidth]{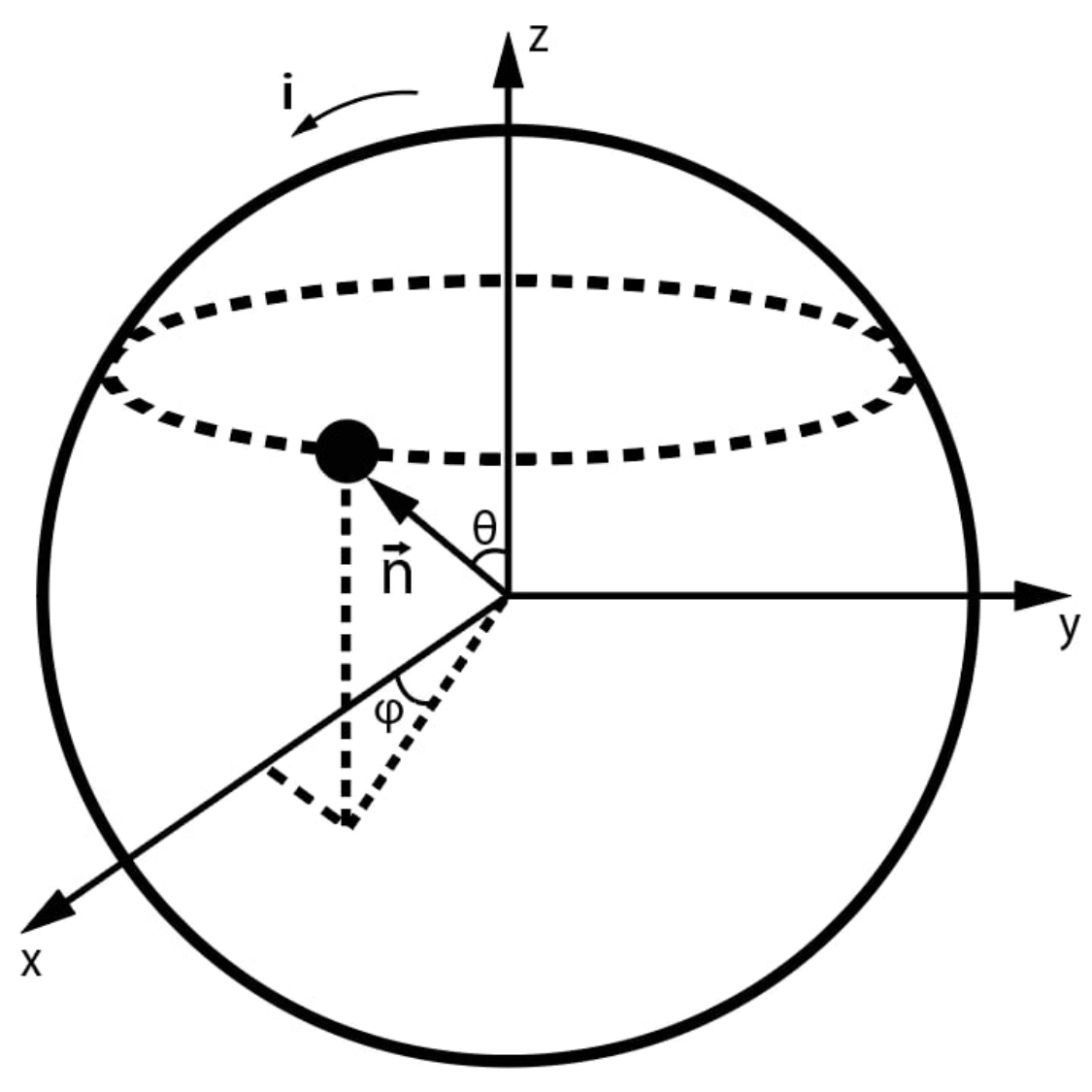}
    \caption{Schematic view of a star with a spot.}
    \label{fig:sphere}
\end{figure}

The observed periodic changes in the radial velocity of the helium lines are related to stellar rotation. For a point on the stellar surface, the change in radial velocity is described by:
\begin{equation}
    V_r^{rot}=v\sin{i}\sin{\varphi}\sin{\theta}, \label{eq:vrot}
\end{equation}
where $v$ represents the linear rotational velocity of the star at the equator, $\varphi$ denotes the longitude, $\theta$ is the angle between the rotation axis and the direction to the spot (latitude), and $i$ is the angle between the rotation axis and the line of sight (see Fig. \ref{fig:sphere}).
This variability model was considered in \citet{2020MNRAS.497.2142M} to determine the latitude $\theta$ of the accretion zone.

However, if the gas moves relative to the surface, the formula (\ref{eq:vrot}) must include an additional term. For a radial motion, this term is expressed as:
\begin{equation}
V_r^{in}=v_{\rm in}(\cos{\varphi}\sin{\theta}\sin{i} + \cos{\theta}\cos{i}),
\end{equation}
where $v_{\rm in}$ represents the velocity of gas inflow. The general case is considered in the appendix.

Thus, in the case of a radial motion, the variation in velocity is the sum of two periodic functions:
\begin{multline}
 V_r^{rot} + V_r^{in} = A\sin{\varphi} + B\cos{\varphi} + C = \\
 = \sqrt{A^2+B^2}\sin{(\varphi+\Delta\varphi)} + C, \label{eq:vrotvin}
\end{multline}
where  $A=v\sin{i}\sin{\theta}$, $B=v_{\rm in}\sin{\theta}\sin{i}$, $C=v_{\rm in}\cos{\theta}\cos{i}$ is the observed mean line velocity, 
$\Delta\varphi$  is the phase shift relative to lines with 
$v_{\rm in}=0:$
\begin{equation}
    \cos\Delta\varphi = \frac{v\sin i}{\sqrt{(v\sin{i})^2+ \left(\frac{C\tg i}{\cos{\theta}}\right)^2}}.
    \label{cosphi}
\end{equation}
Emission lines of metals can be selected as ``stationary'' lines, which should form in the deepest layers of the hot spot, as confirmed by our measurements (Table\,\ref{tab:vHe}).

\begin{figure}[h]
    \centering
    \includegraphics[width=\columnwidth]{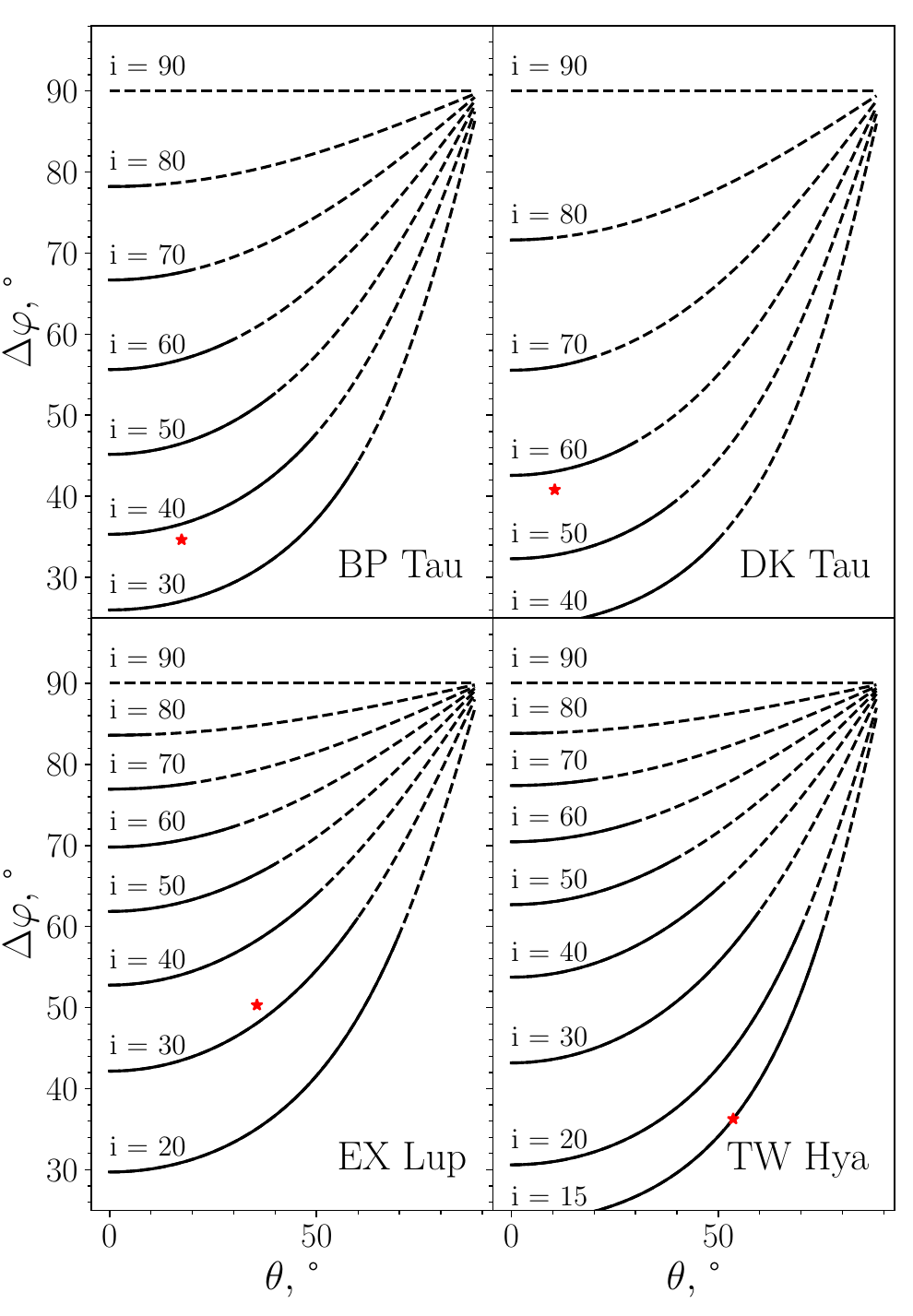}
\caption{Possible values of phase shifts of the He\,II 4686 {\AA} line for all the stars. The dashed lines indicate the regions where the spot is not visible for a given inclination of the stellar rotation axis to the line of sight $i$. The red dots indicate the phase shifts for the star parameters $i, v\sin{i}$ (Table\,\ref{tab:starpar}), and latitude $\theta$ calculated using the formula \ref{eq:vrotvin}.}
\label{fig:phi_theta}
\end{figure}

In Fig.\,\ref{fig:phi_theta} the red dots represent the phase shifts $\Delta\varphi$ of the He\,II radial velocity curve for the parameters of stars $i$, $v\sin{i}$ from the Table\,\ref{tab:starpar}, and the latitude of the spot $\theta$ calculated using the formula \ref{eq:vrotvin} that considers gas inflow. The latitude of the spot is determined by the amplitude of the velocity variation. Formulas \ref{eq:vrot} and \ref{eq:vrotvin} are written for a point spot, but the extent of the spot in longitude can reduce the velocity amplitude, thereby decreasing $\theta$.

The parameter $i$ can also be determined unreliably, therefore we present in the figures the phase shifts that should be observed for any values of $i$ and $\theta$ for the observed $v\sin{i}$ and $C$ for the He\,II 4686  line (Eq. \ref{cosphi}). Similar graphs can be calculated for the He\,I lines. These graphs indicate that, for a fixed $i,$ the phase shift cannot be less than a certain limit, which is reached when $\theta=0\degree.$ For all lines, these limit values at $i$ from the Table\,\ref{tab:starpar} are provided in the Table\,\ref{tab:limits}.

\begin{table}[h]
\caption{Lower limits on phase shifts for observed helium line velocities}
\label{tab:limits}
\centering
\begin{tabular}{c c c c c}
\hline\hline
Line &BP\,Tau&DK\,Tau&EX\,Lup&TW\,Hya\\
 \hline
He\,I\,5876 & $19 \degree$ & $28 \degree$ & $26 \degree$ & $15 \degree$ \\
He\,I\,6678 & $ 9 \degree$ & $17 \degree$ & $17 \degree$ & $  15 \degree$\\
He\,II\,4686 & $33 \degree$ & $40 \degree$ & $44 \degree$ & $ 24 \degree$ \\
\hline
\end{tabular}
\end{table}

\subsection{Observed phase shifts}\label{DeltaPhi}

\begin{figure}[h]
    \centering    \includegraphics[width=0.95\columnwidth]{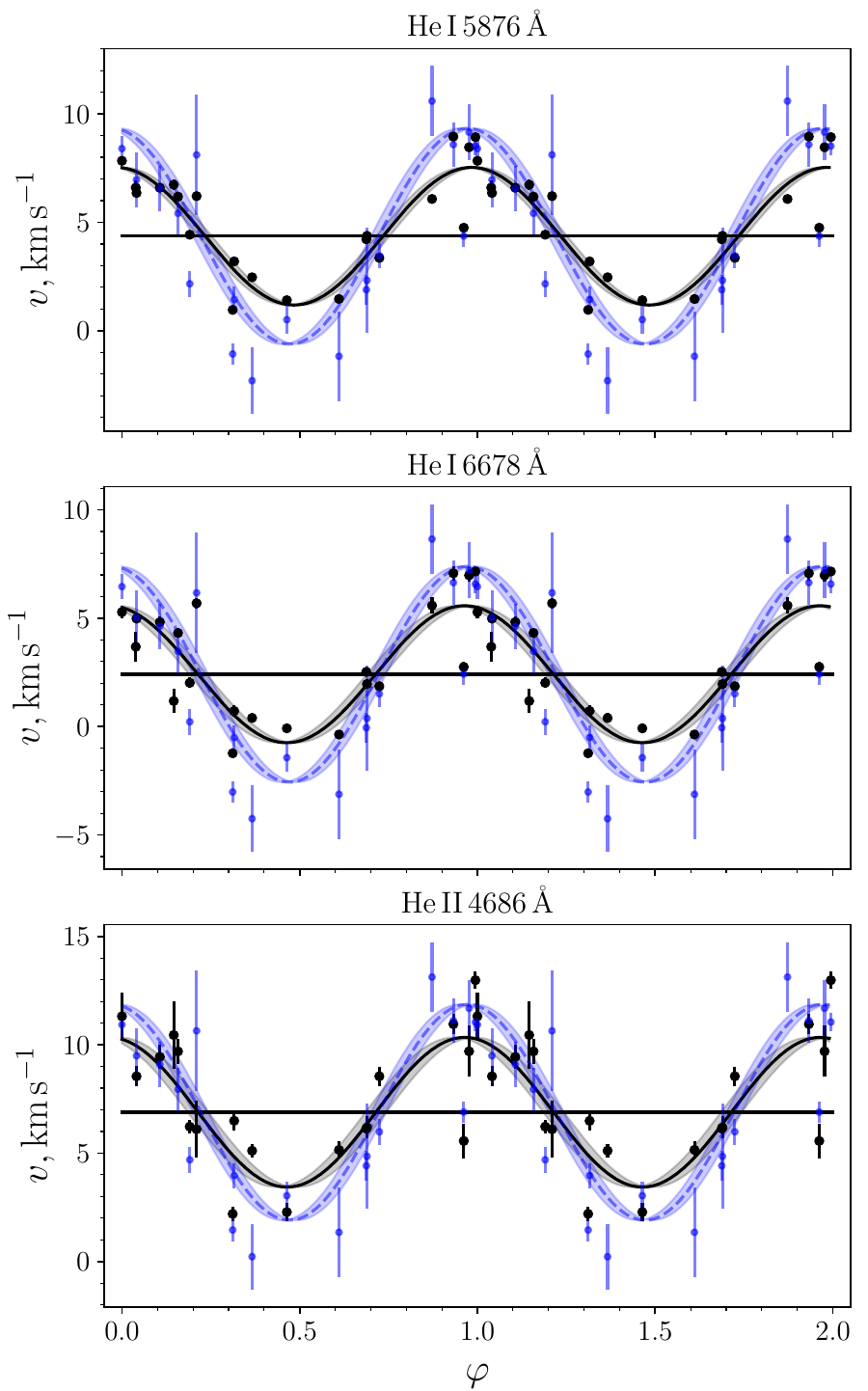}
    \caption{Phase curves of radial velocities of helium lines (black curves and dots) for DK Tau at a period of P = 9.3902 days are compared with curves for metal lines (blue dashed curves and dots). For ease of comparison, the average velocity of the metals is shifted to the average velocity of the corresponding lines.}
    \label{fig:sin3DK}
\end{figure}

\begin{figure}[h]
    \centering
   \includegraphics[width=0.95\columnwidth]{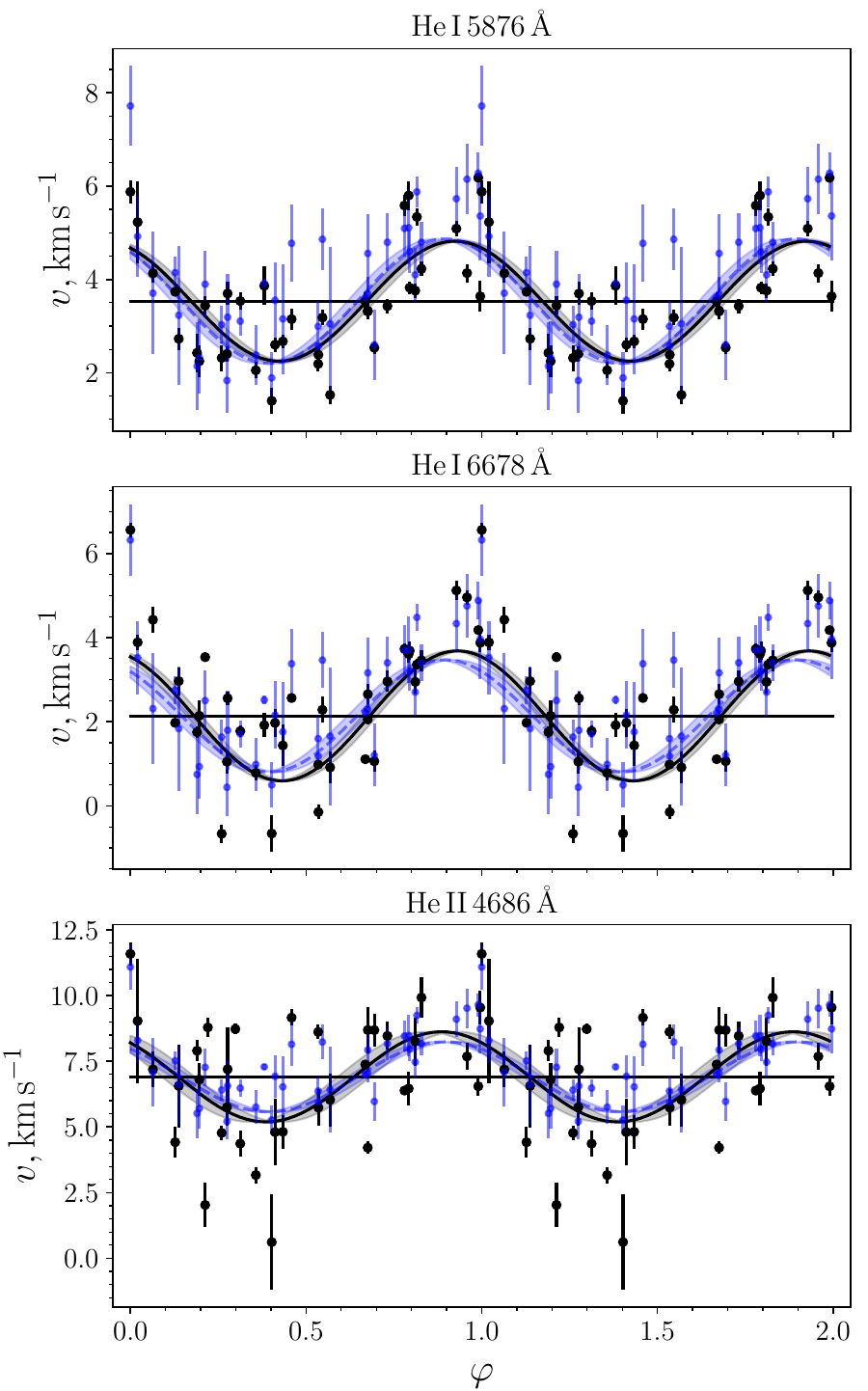}
    \caption{The phase curves of the radial velocities of the helium lines (black curves and dots) for EX Lup are compared with those of the metal lines (blue dashed curves and dots). For ease of comparison, the average velocity of the metals is shifted to the average velocity of the corresponding lines.}
    \label{fig:sin3EX}
\end{figure}

Figures \ref{fig:sin3DK} and \ref{fig:sin3EX} show the radial velocity curves of helium lines are compared with metal lines for DK Tau and EX Lup. The approximating sinusoids were determined using the least-squares. The Table\,\ref{tab:phi} presents the corresponding values of the observed phase shifts relative to the metal lines.  It is evident that within the uncertainties, the phase shifts are equal to zero. For DK Tau, the same result is obtained for other possible period values, such as $P = 8.1887$ days. This is because when the values of two quantities change in phase, they correlate, and this correlation remains consistent for any period value.

\begin{table}
\caption{Observed phase shifts of emission lines relative to metal lines.} \label{tab:phi}
\centering
\begin{tabular}{c c c}
\hline\hline
 Star & Line  & $\Delta\varphi, \degree $ \\
 \hline
DK Tau & He\,I 5876 &  $4.4 \pm 6.8 $ \\
& He\,I 6678  &  $-2.9 \pm 8.9 $ \\
& He\,II 4686  &    $-2.6 \pm  10.5$ \\
& Metals  &    $0 \pm 7.7 $ \\
\hline
EX Lup & He\,I 5876  &  $9.9 \pm 8.8 $ \\
& He\,I 6678  & $12.9 \pm 7.8$   \\
& He\,II 4686  & $-3.3 \pm 16.8$    \\
& Metals  &    $0 \pm 9.9 $ \\
\hline
TW Hya & He\,I 5876  &  $9.8 \pm 18.0$     \\
& He\,I 6678  & $5.7 \pm 10.2$   \\
& Metals  &    $0 \pm 10.8 $ \\
\hline
\end{tabular}
\end{table}

\begin{figure}[h]
    \centering
    \includegraphics[width=\columnwidth]{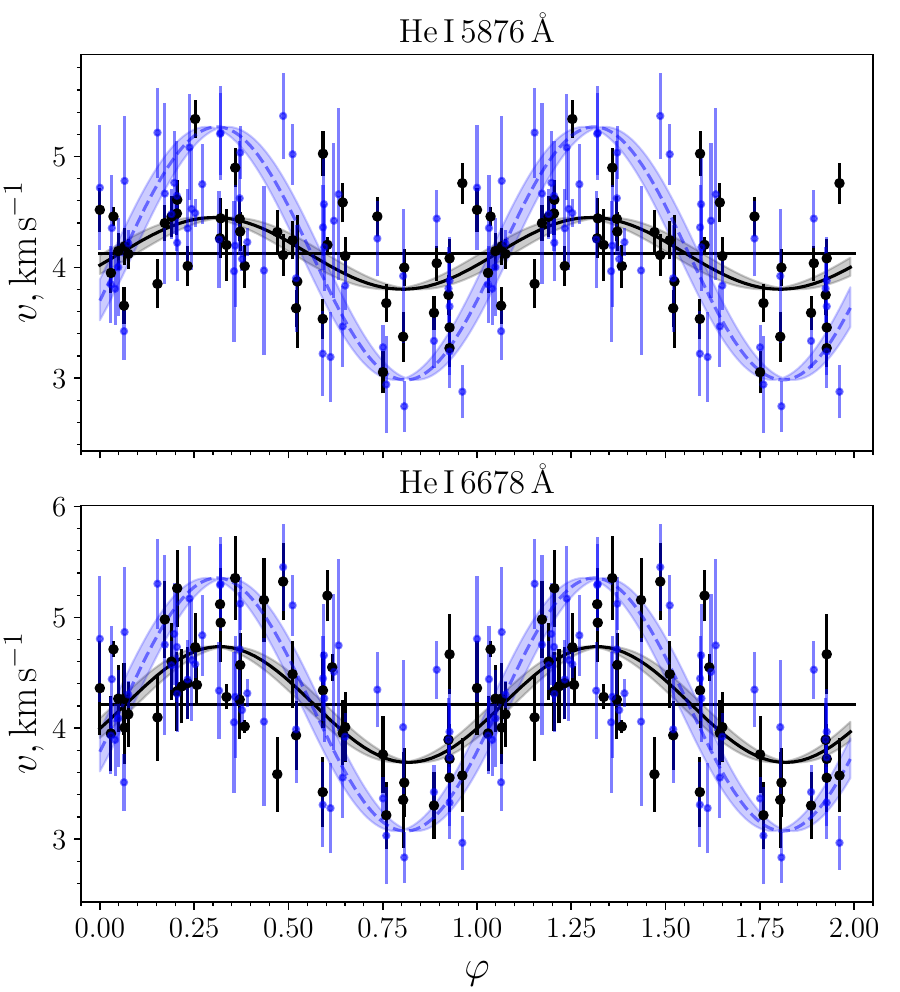}
    \caption{The phase curves of the radial velocities of the He\,I 5876\,\AA \: and He\,I 6678\,{\AA} lines (black curves and dots) of TW Hya are compared with the curves for the metal lines (blue dashed curves and dots). For ease of comparison, the average velocity of the metals is shifted to the average velocity of the corresponding lines.}
    \label{fig:sin2TW}
\end{figure}

Fig. \ref{fig:sin2TW} shows the phase curves of the radial velocities of neutral helium lines compared with metal lines for TW Hya. As with DK Tau and EX Lup, the observed phase shifts between the emission lines in TW Hya are zero, the corresponding values are listed in the Table\,\ref{tab:phi}. Measurements of the He\,II 4686\,{\AA} line velocity in TW Hya do not exhibit periodic variability. This may be due to the presence of a strong broad line component, making it difficult to accurately determine the position of the center of the narrow component.

\begin{figure}[ht]
    \centering
    \includegraphics[width=\columnwidth]{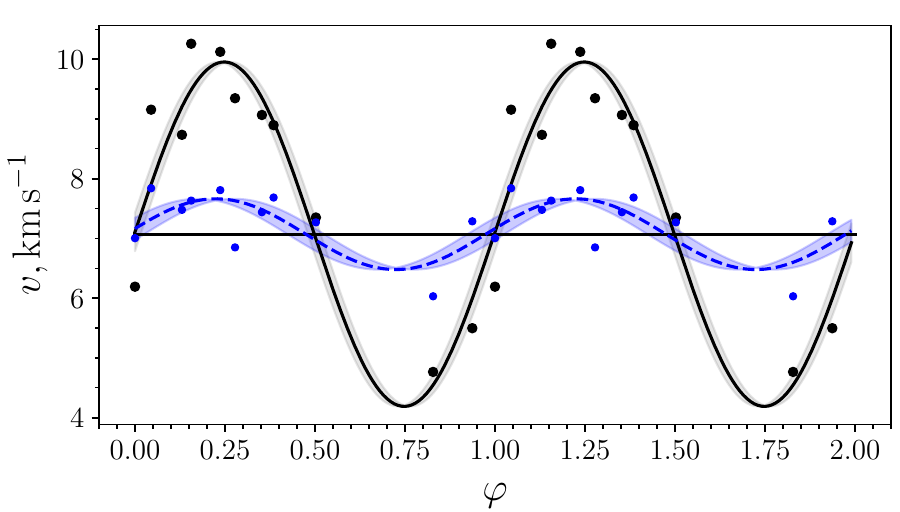}
    \caption{The radial velocity curve of He\,II 4686\,{\AA} (black curve) and He\,I 5876\,{\AA} (blue dashed curve) for the star BP Tau at the epoch $\rm{MJD_0} = 56667.41252.$}
    \label{fig:phiBP}
\end{figure}

In the case of BP Tau, the measurements of the metal line velocities were not accurate enough to determine the phase shifts. Therefore, Fig.\,\ref{fig:phiBP} shows a comparison of the radial velocity curves of the He\,II 4686\,{\AA} and He\,I 5876\,{\AA} lines at a period of P = 8.6681 days during the third observing season (Table\,\ref{tab:observations}), as periodic variability of the lines is observed only in this season. The phase shift between the curves $\Delta\varphi = -8 \degree \pm 14 \degree$ indicates that ionized helium changes in phase with neutral helium within the uncertainties. It is observed that the narrow component of the He\,II 4686\,{\AA} line exhibits a larger amplitude than other emission lines of BP Tau. This can be attributed to the greater longitudinal extent of the formation region of the narrow He\,I component compared to that of the He\,II component (see fig. 6 in \citealt{2024arXiv240405420S}).

Using the formula \ref{cosphi}, we present in Fig.\,\ref{fig:phi_theta} the expected phase shifts, if the observed line shifts are interpreted as Doppler shifts caused by gas inflow. Using the same formula, we can determine the inflow velocities that correspond to the observed shifts. These velocities depend on the latitude of the spot and, similar to the phase shifts, an upper limit can be determined for them, as shown in the Table\,\ref{tab:limitsv} for different lines. Since the metal lines in the case of BP Tau do not exhibit periodic variability, we calculated the maximum possible velocity of the He\,II line relative to He\,I, which was determined to be $v_{\rm He\,II}-v_{\rm He\,I} = 1.7$ \kms.

\begin{table}[h]
\caption{Upper limits of the $C$ line shifts for the observed phase shifts.}
\label{tab:limitsv}
\centering
\begin{tabular}{ c c c c}
\hline\hline
 Line  &DK\,Tau &EX\,Lup &TW\,Hya\\
 \hline
  & & $v$, \kms & \\
He\,I\,5876 & $0.6$ & $1.2$ & $2.6$ \\
He\,I\,6678 & $0.4$ & $1.6$ & $1.5$ \\
He\,II\,4686 & $0.4$ & $0.4$ & --  \\
\hline
\end{tabular}
\end{table}

Therefore, within the uncertainties $\sim10\degree$, all emission lines vary without the phase shifts which would be expected if the gas in their formation region had a velocity relative to the surface of the star.

\section{Interpretation of the observed line shifts}\label{sect:disc}

Our measurements confirm the presence of a shift in helium emission lines by $2-7$ {\kms} relative to the average velocity of absorption lines, which is comparable to the value of $v \sin i$ of the stars. If these shifts are caused by gas accreting onto the star, then significant phase shifts should arise between the radial velocity curves of the different lines. However, all emission lines vary in phase within the measurement accuracy $\sim 10\degree,$ indicating no motion of the gas relative to the stellar surface. Below, we discuss potential causes of line shifts that are unrelated to the Doppler effect.

\subsection{He\,I}

A narrow component of neutral helium originates at the base of the accretion column, where the density reaches values of $\sim10^{15}$ cm$^{-3}$ \citep{2018MNRAS.475.4367D}. At such densities, the line may shift due to the Stark effect. The relationship of this shift to electron density is obtained from the tables \cite{1990A&AS...82..519D}. For the He\,I 6678\,{\AA} line at a density of $N = 10^{14}$ cm$^{-3}$, a redshift of $4.91 \times 10^{-3}${\AA} corresponds to a velocity of $v = 0.2204$ \kms. The observed He\,I 6678\,{\AA}  velocity of $\sim 2$ \kms, based on the linear dependence approximation, corresponds to a density of $\sim 10^{15}$ cm$^{-3}$ for all stars. This is completely consistent with the model values in the region where the narrow components of neutral helium lines originate. Therefore, the He\,I 6678\,{\AA} shift can be completely explained by the Stark effect, rather than by the gas inflow into the spot.

The shift for the He\,I 5876\,{\AA} line cannot be explained similarly, as the shift due to the Stark effect for this line is $-9.86 \times 10^{-4}$\AA, meaning a shift towards the blue side. 
However, the He\,I 5876\,{\AA} line contains several fine structure components that saturate subsequently. Therefore, the position of the line center depends on the optical thickness, as the profile of the summed line can be expressed as
$$f_{\lambda} \sim 1-e^{-\tau_{\lambda}},$$
where
$$
    \tau_{\lambda} = \tau \frac{\sum_i (gf)_i e^{\frac{-(\lambda-\lambda_i)^2}{2\sigma^2}}}{\sum_i (gf)_i}.
$$
Here, $\lambda_i$ and $(gf)_i$ represent the central wavelengths and oscillator strengths of each He\,I 5876\,{\AA} line component, respectively, while $\sigma \approx 0.1$\,{\AA} determines the width of the components.

\begin{figure*}[ht]
    \centering
    \includegraphics[width=\textwidth]{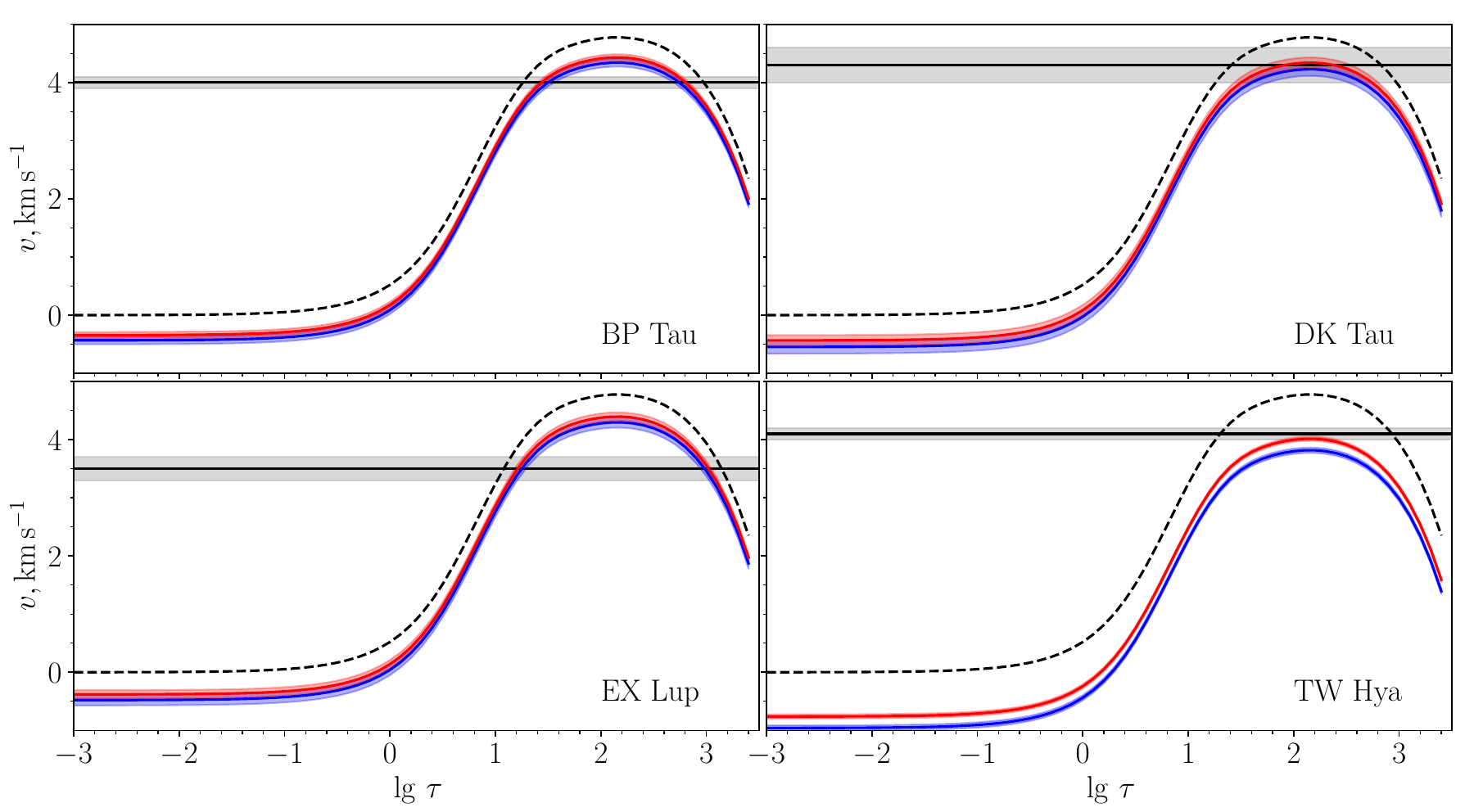}
    \caption{The shift in {\kms} of the He\,I 5876\,{\AA} line as a function of the logarithm of the optical depth $\log{\tau}.$ Red lines correspond to a temperature of 20\,000 K, while blue lines correspond to 10\,000 K. The curves are calculated based on the observed densities derived from the He\,I 6678 \AA. The dashed line represents the He\,I 5876\,{\AA} velocity without considering the shift due to the Stark effect. The horizontal line is the observed velocity of the He\,I 5876\,{\AA} line.}
    \label{fig:stark}
\end{figure*}

The Fig.\,\ref{fig:stark} demonstrates the variation in the position of the He\,I 5876\,{\AA} line center as a function of optical thickness. The graphs are calculated considering the Stark effect at temperatures of $10^4 -2\times10^4$ K and densities that correspond to the observed velocity of the He\,I 6678\,{\AA} line relative to the  $gf$-averaged  wavelength of the He\,I 5876\,{\AA} line components. The region where the observed velocity of He\,I 5876 {\AA} intersects with the calculated shift velocity due to line components, taking into account the Stark effect, is achieved at $\tau \gg 1.$

Thus, the shift of the He\,I 5876\,{\AA} line can be explained if this line is formed in a gas with a density of $10^{15}$ cm$^{-3}$ at a temperature of $10^4 -2\times10^4$ K and at a large optical depth, which is fully consistent with the hotspot models.

\subsection{He\,II}

The provided explanation for the shifts observed in neutral helium lines is not applicable to ionized helium. If the He\,II 4686{\AA} line were optically thick, the shift caused by fine structure of the line would be $\sim -5$ \kms, moving towards the blue side of the spectrum rather than the red side. However, models predict that this line is optically thin, thus this effect is negligible. The Stark effect does not shift the central wavelength of hydrogen-like ions. We cannot identify a specific cause for the shifts in He\,II line velocities. However, the absence of phase shifts between all lines suggests that these shifts are not related to gas motion.
The absence of phase shifts can be observed at a tangential fall onto the star (toroidal velocity $v_{\varphi}\neq0$, see Appendix). However, this type of fall does not result in a shift in the line center, it is possible only with a non-zero radial $(v_{r})$ and/or poloidal $(v_{\theta})$ component of the gas flow velocity. However, in this scenario, phase shifts would occur. 
Velocity shifts may be associated with the distortion of the line profile, either due to the presence of an emission component shifted by several {\kms} to the red side without experiencing rotational modulation, or due to the presence of an absorption component on the blue side of the profile, which also does not experience rotational modulation. These components are not necessarily linked to plasma emitting or absorbing in the He\,II line but may result from the subtraction of stellar absorption lines. Our procedure assumes that the relative depths of the absorption lines in the star and the template spectrum are identical, and that the veiling near the He\,II line remains constant. It cannot be dismissed that similar effects also significantly influence the results of measuring the velocity of the He\,I lines. However, in this case, the correspondence of the density and optical thickness of the He\,I lines to theoretical expectations appears to be a surprising coincidence.

\section{Conclusion}

In this work the velocities of neutral and ionized helium emission lines, as well as the velocities of metal emission lines in the spectra of T Tauri stars were measured. These measurements were based on spectral monitoring of four stars: BP Tau, DK Tau, EX Lup, and TW Hya.

In the cases of DK Tau, EX Lup, and TW Hya, the spectral variability remained consistent throughout the entire observation interval. This indicates that the position and geometry of the accretion spot did not change significantly during this period. Conversely, BP Tau exhibited a different variability pattern, with regular variability observed only during the third observational season (2014).

The average velocities obtained for the He\,I and He\,II lines of the stars are non-zero (Table\,\ref{tab:vHe}) and may suggest the presence of gas movement in the atmosphere. However, such motion should result of a phase shift in the radial velocity curves of the He\,I and He\,II lines relative to the metal lines. Measurements indicate that these phase shifts are absent. 
This indicates that the observed shift of the lines is not associated with the movement of gas in the post-shock. It cannot be ruled out that the shifts may be caused by the movement of another gas, which is not associated with the hotspot, does not exhibit rotational modulation, but is superimposed on the profile from the red side, thereby causing a general shift. However, the observed shifts for neutral helium can be completely explained by shifts caused by the large optical thickness and the Stark effect. The required values of thickness and density are completely consistent with theoretical models for the base of the accretion column \citep{2018MNRAS.475.4367D}.

\acknowledgements
We thank the referees for their thorough review of the article and their valuable comments. The work was supported by the Russian Scientific Foundation (project 23-12-00092).

\label{lastpage}

\bibliographystyle{mypazh}
\bibliography{dip}

\section{Appendix. The radial velocity of the line in the case of gas inflow at an arbitrary angle to the spot.} \label{sect:appendix}

When matter falls onto the surface of a star at an arbitrary angle, the velocity vector of the falling gas can be decomposed into poloidal ($v_{\rm in}^{\theta}$), toroidal $(v_{\rm in}^{\phi})$, and radial $(v_{\rm in}^{r})$ components. Then, the variation in the observed velocity can be expressed as
\begin{multline}
    v = (v\sin\theta - v_{\rm in}^{\varphi})\sin\varphi\sin{i} + \\
    + (v_{\rm in}^r\sin\theta - v_{\rm in}^{\theta}\cos\theta)\cos\varphi\sin{i} + \\
    + (v_{\rm in}^r\cos\theta + v_{\rm in}^{\theta}\sin\theta)\cos i.
\end{multline}

In this case, the phase shift depends on the velocity components of the falling gas and cannot be expressed through observable quantities. The toroidal component alters the amplitude of the velocity variability. The presence of the poloidal component can either increase or decrease the amplitude, phase shift, and mean velocity shift. Nevertheless, the fall of matter at a significant angle does not align with the generally accepted model of accretion. Therefore, we consider the fall of matter to be nearly perpendicular.

\end{document}